# Polynomial algorithm for exact calculation of partition function for binary spin model on planar graphs

Ya. M. Karandashev, M. Yu. Malsagov

**Abstract**. In this paper we propose and realize (the code is publicly available at https://github.com/Thrawn1985/2D-Partition-Function) an algorithm for exact calculation of partition function for planar graph models with binary variables. The complexity of the algorithm is O(N^2). Experiments shows good agreement with Onsager's analytical solution for the two-dimensional Ising model of infinite size.

## 1. Introduction

The paper deals with an algorithm that permits us to find the partition function for a set of interacting spins at lattice nodes, given the spins take either of two values (+1 or -1) and their interaction energy is governed by a quadratic function as in the Ising model.

The calculation of the partition function plays an important role in physics, chemistry, computer vision and machine learning. The use of graph models usually involves computation of two quantities: a posterior probability maximum estimate and marginal distributions. The calculation of the latter is closely related to the determination of the partition function [1]. The problem is to find some statistical properties (e.g. marginal probabilities), given a particular set of random variables in a certain graph model.

It is known that there are very few problems for which the partition function can be calculated exactly. In particular, these are problems using planar graphs (two-dimensional grids), tree graphs, or general type graphs of small sizes. When a problem has hundreds or thousands dimensions, it becomes almost insolvable due to its exponential complexity. In that case the use of rough heuristic methods is the only possible approach. However, it is useful to have exact methods at hand (at least, for a limited range of problems) to develop heuristic methods for approximate calculations of partition functions.

The approximate method based on the use of trees and known as the tree reweighting (TRW) method was offered in papers [2, 3]. The TRW algorithm is meant for the approximate calculation of the most probable configuration of hidden variables in cyclic graph models of the general type.

Paper [4] considers another class of solvable models – planar graphs. A graph is called the planar graph if it can be drawn on a plane without its branches intersecting. It was discovered that in a special type of Ising model [5] with spins {-1, +1} and pair-wise interaction the calculation of the partition function is polynomially reduced to the computation of the matrix determinant [6 – 9]. The current work is dedicated to the realization of these methods. It is worth noticing that significantly more accurate algorithms haven't been suggested since the sixties. The same algorithms are also used in machine learning [10, 11].

Section 2 of the paper offers an algorithm based on the works by P. Kasteleyn and M. Fisher. Section 3 gives the experimental results and algorithm efficiency estimations and compares the algorithm with Onsager's analytic solution for the large dimensionality limit.

## 2. The algorithm

The calculation of the partition function on a planar graph amounts to finding the number of perfect matchings adjusted for their weights by using linear algebra methods as was suggested by P. Kasteleyn and M. Fisher in 1961. Some portions of the algorithm were borrowed from book [12] and articles [13, 14].

Let there be a planar graph $G$ for which the partition function should be found. In short, the algorithm includes the following steps:

1. Dual graph $D$ is built for initial planar graph $G$.

2. The nodes of graph $D$ of degree greater than 2 are unfolded in a planarity-retaining manner to produce extended graph $R$ with coupling matrix $W = \{w_{ij}\}$.
3. Skew-symmetric matrix $B = \{b_{ij}\}$ corresponding to the Pfaff orientation of graph $R$ is constructed.
4. The sought-for partition function is equal to the Pfaffian of matrix $A = \{a_{ij} = b_{ij}w_{ij}\}$, which, in turn, is the square root of the determinant, i.e. $Z = \text{Pf}(A) = \sqrt{\det A}$.

Steps 3-4 are known as the FKT (Fisher-Kasteleyn-Temperley) algorithm. Though each of the four steps of the algorithm is known, for first thing there haven't been so far any consistent description of all four steps solving the partition function problem, for another there haven't been their realization except for [10].

Below thorough consideration of each of the four steps is given.

## 2.1. Dual graph construction

Let there be original graph $G$ with coupling matrix $J_{ij}$. Let each node hold a particle whose spin can take either of two values $s_i = \pm 1$. We recognize these values as the node states. It is necessary to compute the partition function over all possible node states:

$$Z = \sum_s e^{-\beta E(s)}, \qquad (1)$$

where $\beta$ is the inverse temperature, and the energy is determined by pair interaction between graph nodes:

$$E(s) = -\sum_{i,j}^{N} J_{ij} s_i s_j. \qquad (2)$$

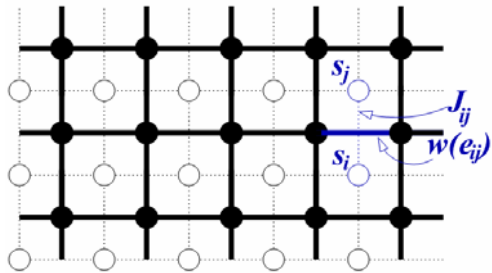

**Fig.1.** The bold lines and black circles designate the dual graph, and the thin lines and white circles stand for the initial graph.

Let $V = \{i : s_i = +1\}$, i.e. $V$ is a collection of nodes whose states are $+1$. Then the expression for the energy can be rewritten in the following manner:

$$E(s) = -\sum_{i,j}^{N} J_{ij} s_i s_j = 2\sum_{i \in V}\sum_{j \notin V} J_{ij} - \sum_{i,j \in V}^{N} J_{ij} - \sum_{i,j \notin V}^{N} J_{ij} = 4\sum_{i \in V}\sum_{j \notin V} J_{ij} - \sum_{i,j}^{N} J_{ij}. \qquad (3)$$

It is seen that the second term $\sum_{i,j}^{N} J_{ij}$ is a constant, and the first term $\sum_{i \in V}\sum_{j \notin V} J_{ij}$ is defined only by node pairs $<i, j>$ with opposite states. It is particularly evident with a planar graph where such node pairs are situated on the boundary of set $V$.

When we deal with planar graph $G$, it is possible to build dual graph $D$ whose nodes are the faces of the original graph and edge weights are determined as follows (see Fig.1):

$$w_{ij} = e^{-4\beta J_{ij}}. \qquad (4)$$

For a dual graph it is possible to say that (see Fig.2) any configuration $s$ of node states in original graph $G$ (Fig.2a) corresponds to a set of Eulerian cycles in dual graph $D$ (Fig.2b). Eulerian cycles are closed curves going along edges $w_{ij}$ and constraining nodes $s_i$ in state +1. It follows that finding partition function (1) amounts to summation over all possible Eulerian subgraphs in dual graph $D$:

$$Z = C \sum_V e^{-4\beta \sum_{i \in V} \sum_{j \notin V} J_{ij}} = C \sum_{\varnothing - Eulerian \atop subgraphs\ of\ D} \prod_{e \in \varnothing} w_e, \qquad (5)$$

where

$$C = \exp\left(\beta \sum_{i,j}^{N} J_{ij}\right). \qquad (6)$$

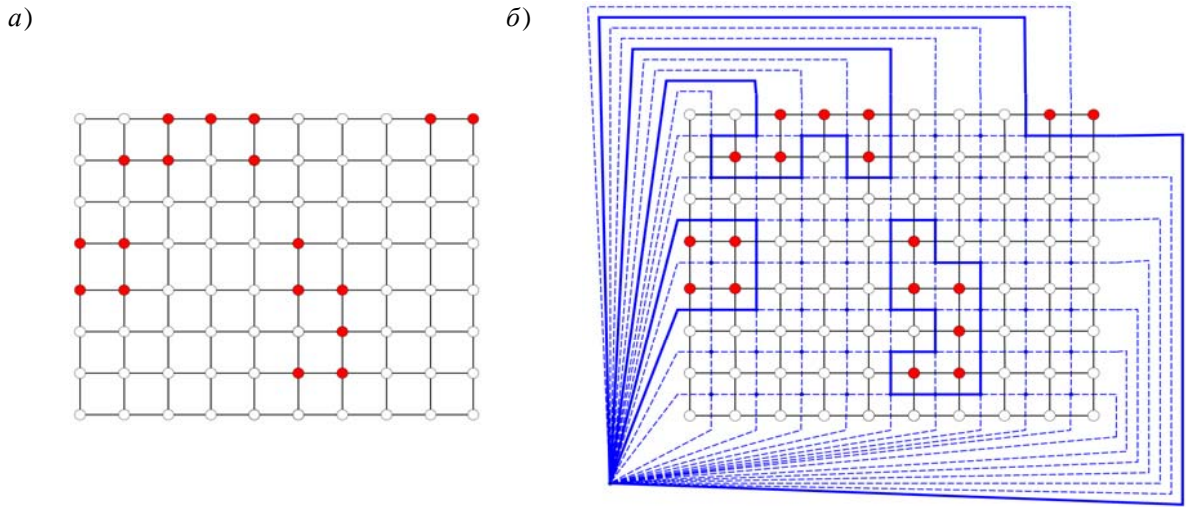

**Fig. 2.** a) The original configuration in which filled circles corresponds to state +1, and blank circles stand for state -1; b) The dual graph (dashed lines and small circles). The Eulerian cycles corresponding to the original configuration are drawn by bold lines.

## 2.2. Unfolding the graph nodes

Let dual graph $D$ be built. We are going to do the two operations:
1. Let us first make all nodes of graph $D$ have degree three by unfolding the nodes of degree greater than three (as shown in Fig.3).
2. Then let us replace each node (which is of degree three now) as shown in Fig.4. This node pattern (see Fig.4b) is chosen because it always allows perfect matching.

The result of the transformation is a new graph, which we call graph $R$. The distinction of the graph is that it remains planar for one thing. For another, it has the even number of nodes and always permits perfect matching. Moreover, it is possible to show that a subgraph of Eulerian cycles in graph $D$ will correspond to each perfect matching in graph $R$ after the reverse operation of node contraction.

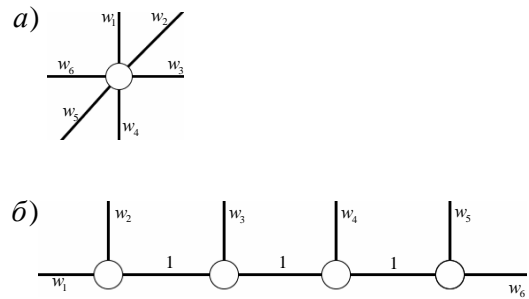

**Fig. 3.** Stage one of graph unfolding. A node of degree greater than three (a) is replaced by several nodes (b).

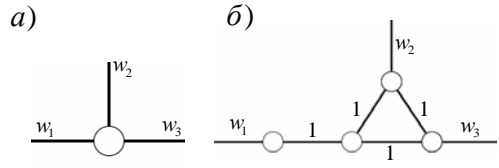

**Fig. 4.** Stage two of graph unfolding. A node (a) is replaced by four nodes (b).

## 2.3. Perfect matching and search for the Pfaffian orientation in the graph

It is known that for any skew-symmetric matrix $A = (a_{ij} : a_{ij} = -a_{ji})$ of size $2n \times 2n$ the Pfaffian can be determined by the formula:

$$\text{pf}(A) = \sum_{\pi} sign \begin{pmatrix} 1 & 2 & \dots & 2n \\ i_1 & j_1 & \dots & j_n \end{pmatrix} a_{i_1 j_1} a_{i_2 j_2} \dots a_{i_n j_n}, \tag{7}$$

where the summation is made over all pair combinations $\pi = \{\{i_1, j_1\},\dots,\{i_n, j_n\}\}$ of set $\{1,\dots,2n\}$, and *sign* denotes the sign of the substitution $\begin{pmatrix} 1 & 2 & \dots & 2n \\ i_1 & j_1 & \dots & j_n \end{pmatrix}$.

Graph R has been undirected so far. Let us assume that in graph $R$ we have chosen a particular edge orientation and designated it as $\vec{R}$. We define matrix $A$ as

$$a_{ij} = b_{ij} w_{ij}, \quad \text{where } b_{ij} = \begin{cases} 1, & \text{if } (i,j) \in e(\vec{R}) \\ -1, & \text{if } (j,i) \in e(\vec{R}) \\ 0, & \text{else} \end{cases} \tag{8}$$

Note that for graph $\vec{R}$ each term of sum (7) belonging to pairing $\pi = \{\{i_1, j_1\},\dots,\{i_n, j_n\}\}$ corresponds to a certain matching. If a particular edge is missing in the matching (i.e. its weight is zero $a_{i_k j_k} = 0$), then the whole product $a_{i_1 j_1} a_{i_2 j_2} \dots a_{i_n j_n}$ is zero. It means that the summation (7) is equivalent to the summation only over available perfect matchings of graph $\vec{R}$. Since each perfect matching in graph $R$ corresponds to a collection of Eulerian cycles in graph $D$ (see previous paragraph), the partition function can be calculated by trying all perfect matchings:

$$Z = C \sum_{\substack{\varnothing - \text{Eulerian} \\ \text{subgraphs in } D}} \prod_{e \in \varnothing} w_e = C \sum_{\substack{PM - \text{perfect} \\ \text{matchings in } R}} \prod_{e \in PM} w_e \geq C \cdot |\text{Pf}(A)|. \tag{9}$$

The inequality in the right side is caused by the fact that the summation in Pfaffian (7) holds terms of opposite signs. On the other hand, if the edge orientation in graph $\vec{R}$ was chosen so that matrix $B = \{b_{ij}\}$ would compensate the signs of permutations $\pi$, then all non-zero addends in sum (7) would appear with the plus sign and we would get

$$Z = C \cdot \text{Pf}(A). \tag{10}$$

The corresponding orientation $B = \{b_{ij}\}$ is called Pfaffian orientation and, what is most important, according to Kasteleyn's theorem it really exists in planar graphs and can be found in polynomial time.

The criterion that can be used to check the Pfaffian orientation is the following theorem:

*If $\vec{R}$ is a connected directed planar graph whose faces (perhaps, except for the infinite face) have an odd number of clockwise directed edges, this directed graph is a Pfaffian graph $\vec{R}$.*

The algorithm of finding the Pfaffian orientation is as follows (see Fig.5). We apply induction on the number of edges in graph $R$. If graph $R$ is a tree, any orientation suits. Let us now assume that it is not a tree and select an edge belonging to the cycle and lying on the boundary of the infinite face. Let $F_0$ be a finite face that holds this edge $e$.

Following the induction, graph $(R-e)$ has the direction where the boundary of each finite face holds an odd number of clockwise directed edges. Let us return edge $e$ to the graph and direct it so that the boundary of face $F_0$ has an odd number of clockwise directed edges. Since all boundaries of finite faces different from $F_0$ have not changed, the resulting orientation of graph $R$ will have necessary properties.

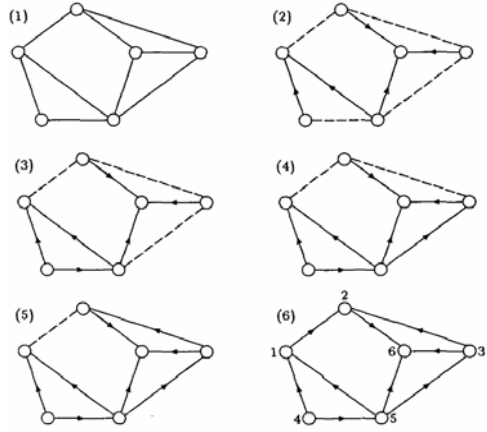

**Fig. 5.** Giving the Pfaffian orientation to a planar graph. (1) The original undirected graph. (2)-(6) The Pfaffian orientation constructing stages.

## 2.4. Calculating the Pfaffian and partition function

After the Pfaffian orientation of graph $\vec{R}$ is found, formula (8) is used to determine matrix $A$. Then it is no problem to compute the partition function:

$$Z = C \cdot \mathrm{Pf}(A) = C \cdot \sqrt{\det(A)}, \qquad (11)$$

because the square of the Pfaffian of a skew-symmetric matrix is equal to the determinant of this matrix.

## 3. Experimental results

The algorithm presented in paragraph 2 was realized and tested on square grids with free boundary conditions and unit coupling weights.

Let us introduce some definitions. The free energy is determined as

$$F = -kT \frac{\ln Z}{N} = -\frac{\ln Z}{\beta N}. \qquad (12)$$

It is more convenient to use quantity $f = \beta F$ which can be called "the normalized free energy" – the term that simplifies expressions for internal energy and thermal capacity:

$$U = \frac{\partial f}{\partial \beta}, \quad C = -\frac{\partial^2 f}{\partial \beta^2}. \qquad (13)$$

In experiments we compared $f$ produced by our algorithm, its first and second derivatives with analytical expressions obtained by Onsanger for the dimensionality approaching the infinity. We used square grids of size $N = n \times n$ when $n = 20, 50, 100$, and $200$.

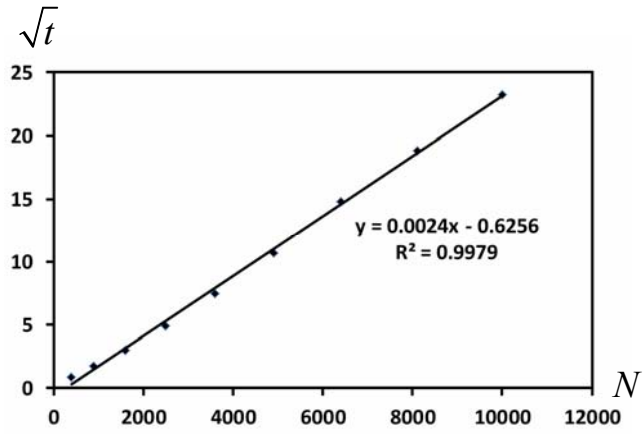

**Fig. 6.** The algorithm operating time $t$ (in seconds).

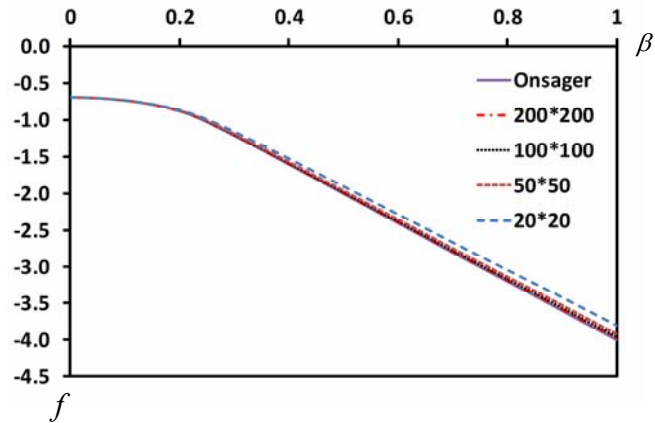

**Fig. 7.** Free energy $f(\beta)$ for grids of different sizes and Onsager's solution for the infinite dimensionality.

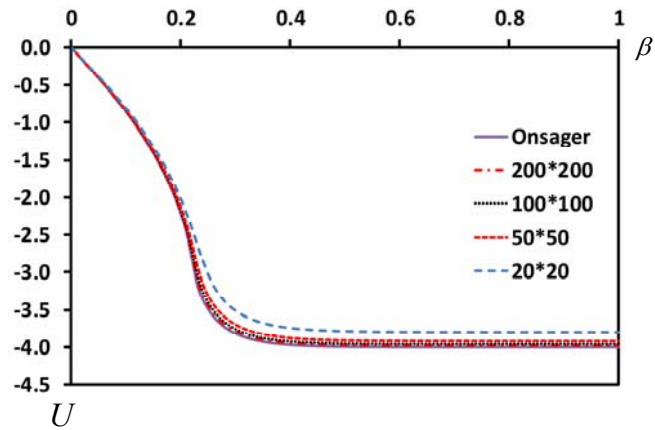

**Fig. 8.** Internal energy $U(\beta)$ for different grids and Onsager's solution for the infinite dimensionality.

Fig.6 shows the dimensionality dependences of the algorithm operating time. To aid the visualization, the square root of time rather than time itself is used for the Y-axis. Thus we get the quadratic algorithm complexity $O(N^2)$ or the fourth power $O(n^4)$ of the linear size of the grid.

Onsager's solution [5] looks as follows

$$f = \lim_{N\to\infty}\frac{Z}{N} = \frac{\ln 2}{2} + \frac{1}{2\pi}\int_0^\pi \ln\left[\operatorname{ch}(4\beta J)\operatorname{ch}(4\beta K) + \frac{\sqrt{1+\chi^2 - 2\chi\cos 2\theta}}{\chi}\right]d\theta, \qquad (14)$$

where

$$\chi = \frac{1}{\text{sh}(4\beta J)\,\text{sh}(4\beta K)}. \tag{15}$$

$J$ and $K$ are the weights of vertical and horizontal couplings in the grid. When $K = J$, the expression for free energy becomes simpler:

$$f = \lim_{N \to \infty} \frac{Z}{N} = \frac{\ln 2}{2} + \ln\left(\text{ch}(4\beta J)\right) + \frac{1}{2\pi} \int_0^{\pi} \ln\left[1 + \sqrt{1 - \lambda^2 \cos^2 \theta}\right] d\theta, \tag{16}$$

where

$$\lambda = \frac{2\,\text{sh}(4\beta J)}{\text{ch}^2(4\beta J)}. \tag{17}$$

Values for free energy, internal energy and thermal capacity are shown in Fig.7 – 9. We computed the first and second derivatives using the finite-difference method and values of free energy at three close points spaced by $d\beta = 10^{-5}$.

As seen from Fig.7-9, the experimental values of free energy and its derivatives approach Osanger's solution with the increasing dimensionality.

With large $\beta$ the partition function is determined mostly by the energy of the global minimum:

$$E_0 = -4n(n-1). \tag{18}$$

That is why the free energy is linearly dependent on $\beta$:

$$f = -4\beta\left(1 - \frac{1}{n}\right). \tag{19}$$

Fig. 7-8 give evidence that the greater the dimensionality, the closer the slope of the line approaches the limit value of -4.

Accordingly, given large $\beta$, the second derivative approaches zero. When $\beta$ takes the critical value $\beta_{cr} \approx 0.22$, the thermal capacity, according to Onsager's solution (14), experiences the break and goes to infinity. It doesn't happen with finite dimensionalities; however, the peak is visible and it becomes more pronounced with dimensionality (Fig.9).

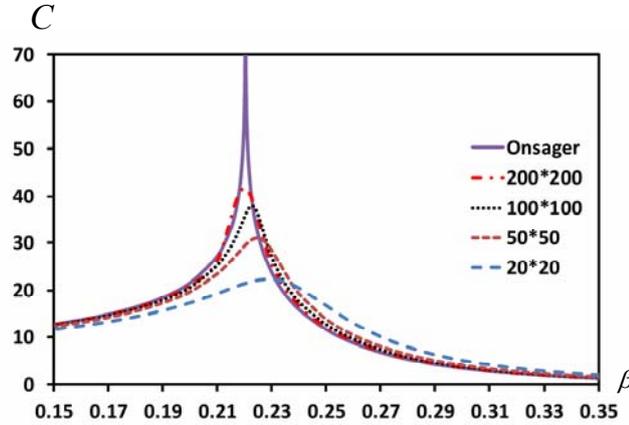

**Fig. 9.** Thermal capacity $C(\beta)$ for different grid sizes and Onsager's solution for the infinite dimensionality. For illustration purpose, the range of $\beta$ in the plot is increased as compared to Fig.7 and 8.

The final experiment was carried out to examine the transfer from the two-dimensional Ising model to single-dimensional one. In the experiment, horizontal coupling weight $K$ varied smoothly from 1 to 0 (see Fig.10), while the vertical coupling weight remained constant $J = 1$. It is seen from Fig.10 that the peaks shift to the right, which corresponds to the increase of the critical temperature $\beta_{cr}$. When $K = 0$, the curve becomes monotonic, the peak disappears and the two-dimensional grid turns into a set of single-dimensional chains for which the phase transition is known to occur at $\beta = \infty$.

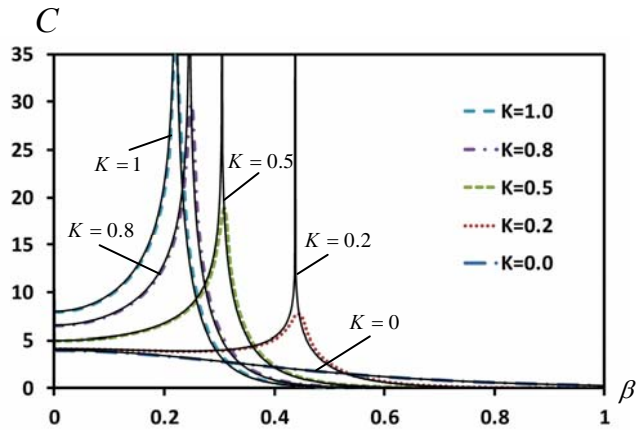

**Fig. 10.** Thermal capacity $C(\beta)$ for different horizontal coupling weights ($K$ varies from 1 to 0) and constant vertical coupling weight ($J = 1$). Solid lines depict Osanger's solution, discontinued curves are generated experimentally for $N = 100$.

## 4. Conclusion

We have offered and realized the algorithm of exact computation of the partition function for two-dimensional binary-variable graph models. The C/C++-based realization of the algorithm uses the Boost and Csparse libraries [15]. The algorithm complexity is $O(N^2)$ and is limited mostly by the time needed for computing the determinant of the sparse matrix.

Further research aims at development of heuristic algorithms for general-type graphs. This kind of algorithm may be used for comparison of approximate results with exact computations.